\documentstyle[12pt,epsfig]{article} 
\def\cg#1#2#3#4#5#6{\mbox{$\langle #1~#2,#3~#4\vert #5~#6\rangle$}}

\def\3j#1#2#3#4#5#6{\mbox{$\left(\begin{array}{ccc} #1 & #2 & #3 \\ #4 & #5 & #6
\end{array} \right)$}}

\def\6j#1#2#3#4#5#6{\mbox{$\left\{\begin{array}{ccc} #1 & #2 & #3 \\ & & \\ #4 &
#5 & #6 \end{array} \right\}$}}

\def\9j#1#2#3#4#5#6#7#8#9{\mbox{$\left\{\begin{array}{ccc} #1 & #2 & #3 \\ #4 &
#5 & #6 \\ #7 & #8 & #9 \end{array} \right\}$}}

\def\oneh{{\textstyle {1\over 2}}} 
 
\def\treh{{\textstyle {3\over 2}}} 
\def\eq{\begin{equation}}
\def\ee{\end{equation}}

\def\eqa{\begin{eqnarray}}

\def\eea{\end{eqnarray}}

\def\bra#1{\mbox{$\langle #1\vert $}}

\def\ket#1{\mbox{$\vert #1\rangle$}}

\def\bbra#1{\mbox{$\langle #1\parallel $}}

\def\kket#1{\mbox{$\parallel #1\rangle$}}

\begin{document}
\pagestyle{empty}
\Huge{\noindent{Istituto\\Nazionale\\Fisica\\Nucleare}}

\vspace{-3.9cm}

\Large{\rightline{Sezione SANIT\`{A}}}
\normalsize{}
\rightline{Istituto Superiore di Sanit\`{a}}
\rightline{Viale Regina Elena 299}
\rightline{I-00161 Roma, Italy}

\vspace{0.65cm}

\rightline{INFN-ISS 97/17}
\rightline{November 1997}

\vspace{0.5cm}
\large
\begin{center} {\large \bf{Nucleon generalized polarizabilities within a
relativistic Constituent Quark Model}} \end {center}

\bigskip
\normalsize
\begin{center} B.  Pasquini$^a$ and G.  Salm\`e$^b$

$^a$Dipartimento di Fisica Nucleare e Teorica, Universit\`a di Pavia, and \\
Istituto Nazionale di Fisica Nucleare, Sezione di Pavia, Pavia, Italy
\\$^b$Istituto Nazionale di Fisica Nucleare, Sezione Sanit\`a, \\ Viale Regina
Elena 299, I-00161 Roma, Italy

\end{center} \begin{abstract} Nucleon generalized polarizabilities are
investigated within a relativistic framework, defining such quantities through a
Lorentz covariant multipole expansion of the amplitude for virtual Compton
scattering.  The key physical ingredients in the calculation of the nucleon
polarizabilities are the Lorentz invariant reduced matrix elements of the
electromagnetic transition current, which can be evaluated from off-energy-shell
helicity amplitudes.  The evolution of the proton paramagnetic polarizability,
$\beta_{\rm {para}}( q)$, as a function of the virtual-photon three-momentum
transfer $ q,$ is explicitly evaluated within a relativistic constituent quark
model by adopting transition form factors obtained in the light-front formalism.
The discussion is focussed on the role played by the effects due to the
relativistic approach and to the transition form factors, derived within
different models.  \end{abstract}

\vskip 1cm


\vskip 1cm

\vspace{1.0cm}
\hrule width5cm
\vspace{.2cm}
\noindent{\footnotesize{  {\bf Phys. Rev. C} in press.}}

\newpage
\pagestyle{plain}
\section{Introduction} One of the major goals of present day researches is to
gain an understanding of the compositeness of the nucleon in terms of the
fundamental quark-gluon dynamics.  An outstanding role in unravelling the
nucleon structure is played by the electromagnetic (em) interactions which
provide clean information on the internal nucleon degrees of freedom.  In
particular, a great deal of attention has been recently devoted to virtual
Compton scattering (VCS) off the nucleon, which is the natural complement to
real Compton scattering and form factor measurements.  VCS will be analyzed
experimentally by means of the $ep\rightarrow e'p'\gamma$ reaction in the
different kinematic domains accessible at Jefferson Lab \cite{[CEBAF]} and at
MIT-Bates~\cite{[MIT]}, while initial data below the pion production threshold
have been taken at MAMI laboratory~\cite{[MAMI]}.

The virtual nature of the initial photon opens the possibility of investigating
the scattering amplitude independently as a function of the momentum and energy
transfers, allowing to access a much greater variety of observables than in real
Compton scattering.  Depending on the photon energy, VCS can be employed to
disentangle different aspects of the nucleon structure:  first, it provides
important insights in the perturbative QCD description of the nucleon wave
function in the hard-scattering regime~\cite{[Farrar]}.  Second, diffractive VCS
has recently attracted new attention in deep inelastic kinematics in order to
get information about some parton distribution inaccessible in standard
inclusive measurements~\cite{[Ji],[Rad]}.  Other energy domains of particular
physical interest are the region of low-lying resonances, where VCS becomes
sensitive to the excitations of the nucleon, and the threshold regime, where one
can measure new em observables which generalize the usual electric ($\alpha$)
and magnetic ($\beta$) polarizabilities.  In particular, at low final photon
energy and arbitrary momentum transfer, it has been shown
~\cite{[Guichon],[D96]} that the nucleon structure information of the VCS
amplitude can be parametrized in terms of a small set of independent generalized
polarizabilities.  Many authors have theoretically investigated the generalized
polarizabilities within the framework of different approaches, such as the non
relativistic constituent quark (CQ) model~\cite{[Guichon]}, effective Lagrangian
models~\cite{[VAN]} and field-theoretical models like the linear sigma
model~\cite{[SIGMA]} and heavy baryon chiral perturbation theory~\cite{[HHKS]}.

Our aim is to study the generalized polarizabilities of the nucleon within a
relativistic scheme based on a Lorentz covariant multipole expansion of the VCS
amplitude, which corresponds to classify the initial and final photon
interaction vertices in terms of Lorentz invariant reduced matrix elements of
the em transition current.  These quantities summarize all the dynamical
information relevant to the VCS process and, in our model, are obtained using
the em transition form factors calculated by adopting a relativistic CQ model,
formulated within the light-front hamiltonian dynamics~\cite{[CPSS]}, and an
off-energy-shell prescription for the baryon em current.  In section 2, we
describe the appropriate formalism to define the generalized polarizabilities,
while in section 3 the results for the evolution of the proton paramagnetic
polarizability, $\beta_{\rm{para}}( q)$, as a function of the virtual-photon
three-momentum transfer $ q,$ is presented.  The sensitivity to both
relativistic effects, introduced by the Lorentz covariance constraint, and
transition form factors, obtained in a relativistic as well as non-relativistic
framework, is discussed.  The conclusions are drawn in the final section.

\section{General formalism} The amplitude for the VCS process $\gamma^*
N\rightarrow\gamma N$ is given by

\eqa T=\varepsilon '^{*}_{\mu}(q')H^{\mu \nu}\varepsilon_{\nu}(q),
\label{1}\medskip \eea where $H^{\mu \nu}$ is the Compton tensor which describes
the interaction between a nucleon and an ingoing virtual photon with
polarization $\varepsilon^{\mu} (q)$ and four-momentum $q^{\mu} \equiv
(\omega,\vec{q})$, followed by the emission of a real final photon with
polarization $\varepsilon '^{\mu}(q')$ and four-momentum $q'^{\mu} \equiv
(\omega'=\vert\vec{q}\,'\vert,\vec{q}\,').$ To leading order in perturbation
theory, $H^{\mu\nu}$ reads \eq H^{\mu\nu}=-i\int{\rm d}^4z\, e^{iq'\cdot
z}\,\bra{\vec p_f\sigma '} T\left(I^\mu(z),I^\nu(0)\right)\ket{\vec p_i\sigma},
\label{2} \ee where $I^{\mu}$ is the hadronic em current and $T(~ , ~)$ denotes
the time-ordered product.  In Eq.~(\ref{2}), $p_{i(f)}^{\mu} \equiv
(E_{i(f)},\vec{p}_{i(f)})$ and $\sigma$ ($\sigma'$) are the four-momentum and
the third component of the spin of the initial (final) nucleon, respectively,
while the normalization of the states is $\bra{\vec p\sigma}\vec
p\,'\sigma'\rangle=(2\pi)^3\,E/m~\delta_{\sigma,\sigma'} \delta(\vec p-\vec
p\,')$.  According to the analysis performed in Ref.~\cite{[Guichon]}, the
Compton tensor can be split into the sum of the nucleon Born term $H^{\mu\nu}_B$
and a residual contribution $H^{\mu\nu}_R$

\eqa H^{\mu\nu}= H^{\mu\nu}_B+H^{\mu\nu}_R, \label{3} \medskip \eea where
$H^{\mu\nu}_B$ includes the gauge invariant contribution from the intermediate
nucleon propagation and from the nucleon-antinucleon excitation and is entirely
determined by the Dirac and Pauli form factors, while the residual part,
$H^{\mu\nu}_R$, provides the new information on the nucleon structure.  The
latter one essentially contains two contributions, the first one from the
nucleon excited states, and the second one from the contact term corresponding
to the seagull diagram derived from the non-relativistic reduction of the em
transition operator.  In our model, the seagull term is not present, since we
assume a relativistic scheme for describing the em interaction and therefore the
residual part reduces to the contribution from the baryon resonant states only.
It is customary (cf.  \cite{[Guichon]}) to parametrize $H^{\mu\nu}_R$ in terms
of generalized polarizabilities obtained by performing a multipole expansion of
the initial and final photon.  This procedure corresponds to classify the
two-step photon-nucleon interaction in terms of the various angular momentum and
parity transfers.  In the low energy VCS limit, at finite $q\equiv|\vec q|$ and
$\omega'\rightarrow 0$, the generalized polarizabilities are defined as \eqa
P^{(\rho 'L',\rho L)S}({q})=\left[\frac{1}{\omega'^{L'} q^L}
H_{R}^{(\rho'L',\rho L)S}(\omega', q)\right]_{\omega'=0}, \label{4}\medskip \eea
where $\rho(\rho')$ indicates the type of the multipole transition ($\rho=0:$
Coulomb; $\rho=1:$ magnetic; $\rho=2:$ electric; $\rho=3$ longitudinal)and
$L(L')$ is the angular momentum of the ingoing (outgoing) photon; $S$ refers to
the spin ($S=1$) and non spin-flip ($S=0$) character of the transition.  In Eq.
(\ref{4}), $H_R^{(\rho'L',\rho L)S}$ are the reduced multipoles introduced in
Ref.~\cite{[Guichon]} and related to $H^{\mu\nu}_R$ as follows \eqa & &
H_R^{(\rho'L',\rho L)S}=\sum_{\sigma,\,\sigma'}
\sum_{M,\,M'}(-1)^{L+M+1/2+\sigma'} \cg{\oneh}{-\sigma'}{\oneh} {\sigma}
{S}{s}\nonumber\\ & &\times\cg{L'}{M'}{L}{-M}{S}{s}{1 \over 4\pi} \int\,{\rm
d}\hat{ q}\,{\rm d}\hat{ q}' V^*_{\mu}(\rho' L' M',\hat q') H^{\mu\nu}_R
V_{\nu}(\rho L M,\hat{ q}), \label{5} \medskip \eea where $V^{\nu}(\rho L
M,\hat{ q})$ are the four-dimensional basis defined in terms of the spherical
harmonics, ${\cal{Y}}_{LM},$ and the vector spherical harmonics,
$\vec{{\cal{Y}}}^{L\ell }_{M},$ by \eqa & & V^{\nu}(0 L M,\hat{ q})\equiv
({\cal{Y}}_{LM}(\hat{q}),\vec{0}), \nonumber\\ & &\nonumber\\ & & V^{\nu}(\rho L
M,\hat{ q})\equiv (0,\sum_{\ell} C^{\rho}_{\ell L} \vec{{\cal{Y}}}^{L\ell
}_{M}(\hat{ q})),~~~~~~~(\rho=1,2,3), \label{6} \medskip \eea with the
coefficients $C^{\rho}_{\ell L}$ given by

\eqa C^1_{\ell\,L}&=&\delta_{\ell,L},\nonumber\\ & &\nonumber\\
C^2_{\ell\,L}&=&\sqrt{\frac{L+1}{2L+1}}\delta_{\ell,L-1}+
\sqrt{\frac{L}{2L+1}}\delta_{\ell,L+1},\nonumber\\ & &\nonumber\\
C^3_{\ell\,L}&=&\sqrt{\frac{L}{2L+1}}\delta_{\ell,L-1}-
\sqrt{\frac{L+1}{2L+1}}\delta_{\ell,L+1}.  \label{7} \medskip \eea In
Eq.~(\ref{5}), by gauge invariance of the Compton tensor, the longitudinal
($\rho=3$) multipole transition may be expressed by the Coulomb ($\rho=0$) one,
while in the Siegert limit, the electric ($\rho=2$) and the Coulomb multipoles
are related by current conservation.  In particular, for the virtual photon
($\rho$) the difference between the electric and Coulomb contribution is given
in terms of mixed polarizabilities, $\hat P^{(\rho'L',L)S},$ whereas for the
real photon ($\rho'$) this term can be neglected to leading order in
$\omega'$~\cite{[Guichon]}.  As a result, by retaining in $H_R^{(\rho'L',\rho
L)S}$ only the linear terms in $\omega'$ and imposing the selection rules of the
angular momentum and parity conservation, one finds only 10 independent
generalized polarizabilities~\cite{[Guichon]}, which are further reduced to 6 by
charge conjugation symmetry~\cite{[D96]}.

In order to achieve a fully consistent relativistic calculation of the nucleon
excitation contribution to the generalized polarizabilities, it is necessary to
develop a proper Lorentz covariant treatement of the multipoles
$H_R^{(\rho'L',\rho L)S}$.  The sum of the contributions of the nucleon excited
states in the direct ($s$) and crossed ($u$) channel leads to the following
definition of the tensor $H^{\mu\nu}_R$ \eqa H^{\mu\nu}_R&=& \sum_{X\neq N,\bar
N} \left(\frac{m_X}{E_X}\, \frac{\bra{N \oneh \sigma';\vec p_f} \,I^{\mu}(0)
\ket {X\, J_X\,\mu_X; \vec{p}_X} \bra {X\, J_X\,\mu_X; \vec{p}_X}
I^{\nu}(0)\ket{N \oneh \sigma; \vec p_i}} {E_f-E_X+\omega'}\right.\nonumber\\ &
&\left.  + \frac{m_X}{E'_X}\, \frac{\bra{N \oneh \sigma';\vec p_f} \, I^{\nu}(0)
\ket {X\, J_X\,\mu_X; \vec{p}\,'_X} \bra {X\, J_X\,\mu_X; \vec{p}\,'_X}
I^{\mu}(0)\ket{N \oneh \sigma; \vec p_i}} {E_i-E'_X-\omega'}\,\right) ,
\label{8} \medskip \eea where the sum covers the whole excitation spectrum of
the nucleon, without the nucleon ($N$) and antinucleon ($\bar{N}$) contribution
taken into account in $H_{B}^{\mu\nu}$, and the on-mass-shell four-momenta of
the intermediate states in the direct and crossed channel are denoted as
$p_X^{\mu} \equiv (E_X=\sqrt{\vec p_X^{\,2}+m_X^2}, \vec p_X=\vec p_f+\vec
q\,')$ and $p'^{\mu}_X \equiv (E'_X=\sqrt{\vec p\,'^{2}_X+m_X^2}, \vec
p\,'_X=\vec p_i-\vec q\, '),$ respectively.

The key physical ingredients in $H^{\mu\nu}_R$ are the matrix elements of the
baryonic em current $I^{\mu}(0)$ between the initial or final nucleon and each
intermediate resonance state.  The multipole decomposition defined in
Eq.~(\ref{5}) allows us, in each matrix element, to separate purely geometrical
factors, as described by the various angular momentum and parity transfers, from
dynamical aspects concerning the baryon structure.  Such a separation can be
achieved by developing a formalism consistent with the constraint of
relativistic covariance~\cite{[KP1]}.  In order to accomplish this, first of
all, we perform the covariant multipole decomposition of the matrix elements
describing the nucleon-resonance transition in the direct channel, then we
generalize the results to the final interaction vertex and to the crossed term
by using the complex conjugation and the crossing symmetry properties,
respectively.

The expansion of the $N-X$ transition matrix element onto the multipole basis of
Eq.~(\ref{6}) reads \eqa & & V_{\mu}(\rho L M;\hat{ q})~\bra {X\, J_X\,\mu_X;
\vec{p}_X} I^{\mu}(0) \ket{N \,\oneh\,\sigma; \vec{p}_i}= \nonumber \\ &
&\nonumber\\ & &\qquad \sum_{\ell\, s}\,\sum_{m\,m_s}\, B^{\rho\,s}_{\ell\, L}\,
\cg{\ell}{m}{s}{m_s}{L}{M}Y^{\ell}_{m}(\hat{ q})\bra{X\,J_X\,\mu_X ;\vec p_X}
I^{s}_{m_s}(0) \ket{N \oneh \sigma;\vec{p}_i}, \label{9} \medskip \eea where the
coefficients $B^{\rho\,s}_{\ell\, L}$ are defined by \eqa B^{0\, s}_{\ell\,
L}=\delta_{s,0},\quad\quad B^{\rho\, s}_{\ell\,
L}=-\delta_{s,1}C^{\rho}_{\ell\,L},\,\,\quad (\rho=1,2,3), \label{10} \medskip
\eea and $I^s_{m_s}$ are the sperical components of the current,
$I^s_{m_s}=(-1)^s I \cdot \xi^s_{m_s}$, with the four-vector basis given by \eqa
\xi^0_0=(1,0,0,0),\qquad \xi^1_{\pm 1}=\mp\textstyle{1 \over \sqrt 2}(0,1,\pm
i,0), \qquad \xi^1_0=(0,0,0,1).  \label{11} \medskip \eea By Lorentz covariance
of the current operator and transformation properties of the baryon states, the
current matrix elements evaluated in a generic frame may be related to the
corresponding ones calculated in the rest frame of the excited resonance by the
well known formula (see e.g.  \cite{[KP2]})

\eqa &&\bra {X\, J_X\,\mu_X; \vec{p}_X} I^{s}_{m_s}(0) \ket{N \,\oneh\,\sigma;
\vec{p}_i} = \sqrt{{E_{{\rm o}\,i} \over E_i}}\,\sum\, \, \Lambda^{s\,\bar s}_
{m_s\,\bar m_{s}}\left({\vec p_X \over m_X}\right)\nonumber\\ & &\nonumber\\ &
&\qquad\times\, D^{1/2}_{\sigma'\sigma}(R_{L}(\vec p_X,\vec p_i)) \bra {X\,
J_X\,\mu_X;\vec{0}} I^{\bar s}_{\bar {m}_{s}}(0) \ket{N \,\oneh\,\sigma';
\vec{p}_{ {\rm o }\,i}}, \label{12} \medskip \eea where, here and in the
following, the sum is understood over repeated indices.  In Eq.~(\ref{12}),
$p^{\mu}_{ {\rm o }\,i}=(E_{ {\rm o }\,i}, \vec{p}_{ {\rm o }\,i})$ is the
nucleon four-momentum in the resonance rest frame.  The spherical components of
the rotationless boost operator, $\Lambda^{s\,\bar s}_{m_s\,\bar{m}_s}(\vec
p_X/m_X),$ can be explicitly expressed through SL(2,C) operators in the form
\eqa \Lambda^{s\,\bar s}_{m_s\,\bar{m}_{s}}\left (\frac{\vec p_X}{m_X}\right)=
(-1)^s~(\xi^{s}_{m_s})^{*}_{\nu} \Lambda^{\nu}_{~\mu} (\xi^{\bar
s}_{\bar{m}_{s}})^{\mu} =(-1)^{\bar{m}_{s}}\oneh \mbox{Tr}
\left\{\sigma^s_{m_s}\, L\left(\frac{\vec p_X}{m_X}\right)\, \sigma^{\bar
s}_{-\bar{m}_{s}}\, L^\dagger\left(\frac{\vec p_X}{m_X}\right)\,\right\},
\label{13} \medskip \eea where $L(\vec p_X/m_X)= (E_X+m_X+\vec{p}_X \cdot
\vec{\sigma})/ ~\sqrt{2m_X( E_X+m_X)}$ and $\sigma^s_{m_s}$ are the spherical
components of the four-vector defined in terms of the Pauli matrices as
$\sigma_{\mu} \equiv (1,\vec{\sigma}).$ The rotationless boost operator maps
into the rest frame of the baryon state, viz.  \eqa (m_X,\vec 0)=
\hat{\Lambda}^{-1}\left({\vec p_X \over m_X}\right) p_X, \qquad\qquad p_{{\rm
o}\,i}=\hat{\Lambda}^{-1}\left({\vec p_X\over m_X}\right) p_i, \label{14}
\medskip \eea while $R_{L}$ is the corresponding Wigner rotation given by \eqa &
& R_{L}(\vec p_X,\vec p_{ i})=L^{-1}\left({\vec p_{{\rm o}\,i}\over m_N}\right)
\,\, L^{-1}\left({\vec p_X\over m_X}\right) \,\, L\left({\vec p_i\over
m_N}\right)= \nonumber \\ & &\qquad{(E_X+m_X)(E_i+m_N)-\vec{p}_X \cdot
\vec{p}_i-i \vec{\sigma} \cdot (\vec{p}_X \times \vec{p}_i) \over
\sqrt{2(E_X+m_X)(E_i+m_N)(E_XE_i-\vec{p}_X \cdot \vec{p}_{\rm i}+m_N m_X)}}\, .
\label{15} \medskip \eea As shown in Refs.~\cite{[KP2],[CK]}, the current matrix
elements in the resonance rest frame can be parametrized in terms of Lorentz
invariant reduced matrix elements according to \eqa & & \bra{X\,J_X\,\mu_X ;\vec
0} I^{s}_{m_s}(0)\ket{N \oneh \sigma;\vec{p}_{{\rm o}\,i}}= \frac{(-1)^s}{\sqrt
2} \sum \bra{\ell\,m \,s\,m_s} J\,M\rangle\nonumber\\ & &\nonumber\\ &
&\qquad\times\bra{\oneh\,\sigma\, J\,M\,}J_X\,\mu_X\rangle
Y^{*}_{\ell,m}(\hat{p}_{{\rm o}\,i})\,\bbra {X\, J_X} I_{[\ell,s,J]}(Q^2)
\kket{N\oneh}, \medskip \label{16} \eea where the reduced matrix elements $\bbra
{X\, J_X}I_{[l,s,J]}(Q^2)\kket{N\oneh}$ 
square four-momentum transfer, $Q^2=-q\cdot q$.  Furthermore they are
constrained by the selection rules dictated by parity and angular momentum
conservation, viz.  \eqa (-1)^{\ell+s}=\Pi\,\Pi_X,\qquad\qquad \vert
J_X-\oneh\vert\leq J\leq J_X+\oneh, \label{17} \medskip \eea with $\Pi$ and
$\Pi_X$ denoting the intrinsic parity of the nucleon and baryon resonance,
respectively.  Finally, by using Eqs.~(\ref{12}) and (\ref{16}) we obtain the
Lorentz covariant multipole decomposition for the matrix element in a generic
frame, at the same $Q^2.$

The corresponding results for the matrix elements describing the $ X - N $
transition can be easily derived by means of the relation \eqa \bra{N \oneh
\sigma; \vec {p}_{\rm i }} I^{s}_{m_s}\ket{X\,J_X\,\mu_X ; \vec p_X}= (-1)^{m_s}
\bra{X\,J_X\,\mu_X ; \vec p_X} I^{s}_{-m_s}\ket{N \oneh \sigma; \vec {p}_i}^*,
\label{18}\medskip \eea which, in terms of the reduced matrix elements, reads
\eqa \bbra {N\,\oneh} I_{[l,s,J]}(Q^2) \kket{X\,J_X}=(-1)^{J_X-\oneh}~\frac{\hat
J_X}{\sqrt 2} \bbra {X\, J_X} I_{[l,s,J]}(Q^2) \kket{N\oneh}, \label{19}\medskip
\eea where $\hat J_X=\sqrt{2~J_X+1}$.  Then, by collecting the results of
Eqs.~(\ref{5}), (\ref{8}), (\ref{9}), (\ref{12}) and (\ref{16}), and after a
straightforward angular momentum algebra, the contribution of the direct term
($s$) to the reduced multipoles takes the form

\eqa & & H^{(\rho'\,L',\,\rho \, L)S}_{ s}= {1\over 32\pi\sqrt{2}\hat S}\sum_X
\sum\,(-1)^{\bar s'+s'+1+{\cal L}+\ell'+S+c+K+ J'} \hat b\hat b'\hat d\hat
d'\hat a\hat a'\hat L\hat L'\hat J \hat{J'}\hat c\hat{J}_X \nonumber\\ &
&\nonumber\\ & &\times \, B^{\rho\,s}_{\ell\,L} B^{\rho'\,s'}_{\ell'\,L'}
\6j{\oneh}{\oneh}{c}{J}{J'}{J_X} \9j{\bar s}{s}{b}{{\cal L}}{\ell }{d}{J}{L}{a}
\9j{\bar s'}{s'}{b'}{{\cal L}'}{\ell'}{d'}{J'}{L'}{a'}
\9j{J}{L}{a}{J'}{L'}{a'}{c}{S}{K} \nonumber\\ & &\nonumber\\ & &\times \int{\rm
d}\hat{ q}\int{\rm d}\hat{ q}'\, \frac {m_X}{E_X}\,\sqrt{\frac{E_{{\rm
o}\,i}E_{{\rm o}\,f}}{E_{i}E_{f} }} \,(-1)^{k} O^{K}_{k}(S,c;\vec p_X,\vec
p_i,\vec p_f) T^{K}_{-k}(\nu,\nu';\hat{q},\hat {p}_{{\rm o}\,i};\hat{q}',
\hat{p}_{{\rm o}\,f};{\vec p_X\over m_X}) \nonumber\\ & &\nonumber\\ & &\times
\frac{\bbra{N\,\oneh}I_{[{\cal L'},\bar s', J']}(0)\kket{X\, J_{X}} \,
\bbra{X\,J_{X}}I_{[{\cal L},\bar s, J]}(Q^2)\kket{N\,\oneh}} {E_f-E_X+\omega'}\,
, \label{20} \eea where the tensor $T^{K}_{k}$ is obtained by coupling the
spherical representation of the Lorentz boost and the spherical harmonics
derived by the multipole expansion of the em field at the first and second
interaction vertex and depends on the indices $\nu \equiv (a,b,s,\bar s,d,{\cal
L},\ell)$ and $\nu' \equiv (a',b',s',\bar s',d',{\cal L'},\ell')$.  The tensor
$O^K_k$ is related to the Wigner rotations $R_L(\vec{p}_X,\vec{p}_{i})$ and
$R_L^\dagger(\vec{p}_X,\vec{p}_{f})$ acting on the initial and final nucleon
state, respectively.  The explicit expressions for the tensors $T^K_k$ and
$O^K_k$ can be found in Appendix.  The analogous expression for the contribution
of the $u$ channel follows from Eq.~(\ref{20}) by using the crossing symmetry
property, corresponding to the $ q^\mu\leftrightarrow -q'^\mu$, $V_{\nu}(\rho L
M, \hat q) \leftrightarrow V_{\mu}^{\ast}(\rho' L' M', \hat q')$, and reads
explicitly

\eqa & & H^{(\rho'\,L',\,\rho \,L)S}_u= {1\over 32\pi\sqrt{2}\hat S} \sum_X
\sum(-1)^{\bar s+s'+1+{\cal L}'+\ell'+L+L'+c+K+ J} \hat b\hat b'\hat d\hat
d'\hat a\hat a'\hat L\hat L'\hat J \hat{J'}\hat c\hat{J}_{X} \nonumber\\ &
&\nonumber\\ & &\times \, B^{\rho\,s}_{\ell\,L}
B^{\rho'\,s'}_{\ell'\,L'}\6j{\oneh}{\oneh}{c}{J'}{J}{J_{X}} \9j{\bar
s}{s}{b}{{\cal L}}{\ell }{d}{J}{L}{a} \9j{\bar s'}{s'}{b'}{{\cal
L}'}{\ell'}{d'}{J'}{L'}{a'} \9j{J'}{L'}{a'}{J}{L}{a}{c}{S}{K} \nonumber\\ &
&\nonumber\\ & &\times \int{\rm d}\hat{ q}\int{\rm d}\hat{ q}'
\frac{m_X}{E'_X}\, \sqrt{E_{{\rm o}\,i}E_{{\rm o}\,f} \over E_{i}E_{f} }
\,(-1)^{k} O^{K}_{k}(S,c;\vec p\,'_X,\vec p_i,\vec p_f)
T^{K}_{-k}(\nu,\nu';\hat{ q},\hat{ p}\,'_{{\rm o}\,f};\hat{ q}\,', \hat{
p}\,'_{{\rm o}\,i};{\vec p\,'_X\over m_X}) \nonumber\\ & & \nonumber\\ & &\times
\frac{ \bbra{N\,\oneh}I_{[{\cal L}, \bar s, J]}(Q^2)\kket{X\,J_{X}} \,
\bbra{X\,J_{X}}I_{[{\cal L'}, \bar s', J']}(0)\kket{N\,\oneh}}
{E_i-E'_X-\omega'}\,.  \label{21} \eea

Eqs.~(\ref{20}) and (\ref{21}) exhibit the required Lorentz covariance, obtained
by the decomposition of the em current operator into spherical tensors which
transform irreducibly under the Lorentz group.  The main result is that the
geometrical and kinematical aspects can be completely factored into the 6-j and
9-j symbols and the Lorentz transformation coefficients given by the tensors
$T^K_k$ and $O^K_k$, while the dynamical content is summarized into the Lorentz
invariant reduced matrix elements of the em current.  The latter ones contain
all the relevant information on the baryon structure and represent the
quantities to be calculated by models.  First of all, it is essential to note
that in the VCS process, below the pion threshold, only off-mass-shell
intermediate baryons are involved; while in the actual calculation an
off-energy-shell prescription for the baryon current has been introduced.  To
this end, we followed the same spirit of the approach proposed by De
Forest~\cite{[DEFOREST]} for the electron-nucleus scattering, namely the
resonance four-momentum is on its mass-shell and only the three-momentum is
conserved, with $\vec{p}_X=\vec p_f+\vec q\,'$ in the direct channel and
$\vec{p}\,'_X=\vec{p}_{i}- \vec{q}\,'$ in the crossed one.  Within such an
approach, we have evaluated off-energy-shell helicity amplitudes defined by \eqa
S^{X}_{1/2}& = & \bra {X\, J_{X}\,\oneh; \vec 0} I^{0}_{0} \ket{N
\,\oneh\,\oneh; -|\vec{p}_{{\rm o} \,i}|~\hat{ z}}, \nonumber\\ & &\nonumber\\
L^{X}_{1/2} & = & \bra {X\, J_{X}\,\oneh; \vec 0} I^{1}_{0} \ket{N
\,\oneh\,\oneh; -|\vec{p}_{{\rm o}\,i}|~\hat{ z}}, \nonumber\\ & &\nonumber\\
A^{X}_{1/2} & = & -\bra {X\, J_{X}\,\oneh; \vec 0} I^{1}_{1} \ket{N \,\oneh\,
-{\textstyle {1\over 2}}; -|\vec {p}_{{\rm o}\,i}| ~\hat{ z}} ,\nonumber\\ &
&\nonumber\\ A^{X}_{3/2}&=& -~\bra {X\, J_{X}\,\textstyle {3\over 2}; \vec
0}I^{1}_{1} \ket{N \,\oneh\,\oneh; -|\vec {p}_{{\rm o}\,i}|~\hat{ z}}.  \medskip
\label{22} \eea By using Eqs.~(\ref{16}) and (\ref{22}), the reduced matrix
elements have been expressed in terms of the off-energy-shell helicity
amplitudes as follows \eqa & &\bbra{X\,J_X} I_{[\ell,s,J]}(Q^2)\kket{N\oneh}=
\Pi_X 2{\sqrt{8\,\pi (2\ell+1)}\over (2J_X+1)}\nonumber\\ & &\times \left[\rule
{0.cm} {.4cm}\bra{\ell\,0\,s\,0} J\,0\rangle \bra{\oneh\,\oneh\,J\,0}J_X\,\oneh
\rangle ( \delta_{s,0} S^X_{1/2} + \delta_{s,1} L^X_{1/2} ) \right.  \nonumber\\
& & \left.  -\bra{\ell\,0 \,s\,1} J\,1\rangle (\bra{\oneh\,\oneh \,J\,1}
J_X\,{\textstyle {3\over 2}} \rangle A^X_{3/2}+ \bra{\oneh\,-\oneh \,J\,1}
J_X\,\oneh \rangle A^X_{1/2}) \rule{0.cm} {.4cm} \right], \label{23}\medskip
\eea where $L^X_{1/2}$ can be eliminated in favour of $S^X_{1/2}$ by current
conservation.

\section{The paramagnetic polarizability }

As a first application of the formalism developed in the previous section, we
analyze the role of both relativistic and transition form factor effects in the
calculation of the dipole magnetic polarizability of the nucleon.  In the center
of mass frame of the final nucleon-photon system and in the limit of $\omega'
\rightarrow 0,$ the polarizabilities depend only on the three-momentum of the
virtual photon.  The $q$ evolution of diagonal magnetic contribution represents
a direct generalization of the $\beta $ polarizability for the virtual photon
case , viz.  \eqa \beta( q)=-\alpha_{QED}\sqrt{\frac{3}{8}} P^{(11,11)0}( q),
\medskip \label{24}\medskip \eea where $\alpha_{QED}\simeq1/137$ and the real
Compton scattering limit is recovered for $ q\rightarrow 0.$

As is well known, the magnetic polarizability consists of a positive
paramagnetic contribution which is largely cancelled by a negative diamagnetic
term due to the polarization of the pion cloud surrounding the nucleon.  Since
we are interested in evaluating only the nucleon resonance excitation
contribution, we limit ourselves to discuss the paramagnetic term $\beta_{\rm
para}( q)$.  Moreover, in our exploratory analysis we consider only the
overwhelming contribution from the $\Delta(1232)$ to the dipole magnetic
excitation of the nucleon \footnote{Within such an approximation the proton and
neutron magnetic polarizabilities are equal since the transition form factors
involved in the calculations are the same.}.

The reduced matrix elements of the $N-\Delta(1232)$ transition current have been
explicitly calculated in the low energy VCS limit by using the off-energy-shell
prescription described in Sect.  2 and the following relativistic current
operator

\eqa I_{\Delta}^{\mu \nu} =\sqrt{{2 \over 3}} ~ \left[G_1^{\Delta}\left(
    Q^2\right ) ~ {\cal{K}}_{1}^{\mu\nu} + G_2^{\Delta}\left( Q^2\right ) ~
    {\cal{K}}_{2}^{\mu\nu} + G_3^{\Delta}\left( Q^2\right )
    ~{\cal{K}}_{3}^{\mu\nu} \right], \medskip \label{25}\medskip \eea where
    $G_i^{\Delta}\left( Q^2\right )$ and ${\cal{K}}^{\mu\nu}_i$ are the
    kinematic-singularity free form factors and the tensors as defined in
    Ref.~\cite{[DEK]}, respectively.  In particular, for the $\Delta(1232)$
    excitation the helicity amplitudes are \eqa A^{\Delta}_{1/2}& =&
    {1\over{\cal E}} \frac{\sqrt 2}{3}\, q\,\left[G_1^{\Delta}\left(Q^2\right)
    (\omega-E_{N}-m_{N}) +G_2^{\Delta}\left( Q^2\right )\omega m_{\Delta} -
    G_3^{\Delta}\left( Q^2\right )Q^2 \right],\nonumber\\ \nonumber\\
    A_{3/2}^{\Delta}& = &{1\over{\cal E}} \sqrt{\frac{2}{3}}\,q\,
    \left[-G_1^{\Delta}\left( Q^2\right )(\omega+E_{N}+m_{N})
    -G_2^{\Delta}\left( Q^2\right ) \omega m_{\Delta} + G_3^{\Delta}\left(
    Q^2\right )Q^2 \right],\nonumber\\ & & \nonumber\\ S_{1/2}^{\Delta}& =&
    {1\over{\cal E}} \frac{2}{3}\,q^2\, \left[G_1^{\Delta}\left( Q^2\right )
    +G_2^{\Delta}\left( Q^2\right )m_{\Delta} + G_3^{\Delta}\left( Q^2\right
    )\omega \right], \label{26} \eea where ${\cal E}=\sqrt{2m_N(E_N+m_N)}$ with
    $E_N=\sqrt{\vec q\,^2+m_N^2}$.  Note that within this off-shell prescription
    the constraint imposed by current conservation in the Siegert limit of the
    helicity amplitude is satisfied, viz.  \eqa \lim_{q\rightarrow 0}\left(
    \sqrt{3}\,A_{1/2}^{\Delta}-A_{3/2}^{\Delta}\right)= \lim_{q\rightarrow
    0}\left( \frac{\omega}{q}\sqrt{6}~S_{1/2}^{\Delta}\right), \label{27}
    \medskip \eea and the corresponding relation for the reduced matrix elements
    (see Eq.~(\ref{23})) in the limit of $q\rightarrow 0$ reads \eqa & &
    \lim_{q\rightarrow 0} \left [ \sqrt{\frac{3}{5}}~\bbra{\Delta\,{\textstyle{3
    \over 2}}}I_{[1,1,2]}(0) \kket{N\,\oneh}+
    \sqrt{\frac{2}{5}}~\bbra{\Delta\,{\textstyle{3 \over 2}}}I_{[3, 1, 2]}(0)
    \kket{N\,\oneh} \right ]= \nonumber \\ & & \lim_{q\rightarrow 0}\left [
    \frac{\omega}{q}\sqrt{\frac{3}{2}} ~\bbra{\Delta\,{\textstyle{3 \over
    2}}}I_{[2, 0, 2]}(0)\kket{N\,\oneh}\right ].  \label{28} \medskip \eea

In order to perform a consistent relativistic calculation, within a Lorentz
covariant formalism as well as a relativistic approach for describing the
internal nucleon dynamics involved in the scattering process, we have evaluated
the magnetic polarizability using the $G_i^{\Delta}$ form factors obtained in
Ref.~\cite{[CPSS]} within the light-front hamiltonian dynamics.  In particular
such form factors have been calculated i) by adopting baryon eigenstates of the
relativized mass operator proposed in Ref.~\cite{[CI]}, which quite well
reproduces the baryon mass spectroscopy, and ii) by using a relativistic CQ
one-body em current which contains Dirac and Pauli CQ form factors as well.

The reduced multipoles have been calculated in the center of mass frame of the
 final photon-nucleon system and as a consequence the boost transformations in
 Eqs.~(\ref{20}) and (\ref{21}) only affect the crossed channel term, where the
 $\Delta(1232)$ is excited with momentum $\vec{p}\,'_{\Delta}=-\vec {q}-\vec
 {q}\,',$ while they reduce to the identity in the direct term, where the
 $\Delta(1232)$ is in its rest frame.  The explicit expression of the $s$
 channel term in the resonance rest frame becomes \eqa & & H^{(\rho'\,L',\,\rho
 \, L)S}_{ s}= {1\over 4\pi\sqrt{2}}\sum_X \sum\,(-1)^{s+s'+1+L'+S} \hat{J}_X \,
 B^{\rho\,s}_{\ell\,L} \,B^{\rho'\,s'}_{\ell'\,L'}
 \6j{\oneh}{\oneh}{S}{L}{L'}{J_X} \nonumber\\ & &\nonumber\\ & &\times
 \frac{\bbra{N\,\oneh}I_{[\ell', s', L']}(0)\kket{X\, J_{X}} \,
 \bbra{X\,J_{X}}I_{[\ell, s, L]}(Q^2)\kket{N\,\oneh}}
 {\sqrt{\omega'^2+m_N^2}-m_X+\omega'}\,, \label{29} \eea while the corresponding
 contribution to the dipole magnetic polarizability reads \eq P^{(11,11)0}_{
 s}={-1\over 4\pi \sqrt{3}}\,\left(\frac{1}{\omega' \,q}\,
 \frac{\bbra{N\,\oneh}I_{[1, 1, 1]}(0)\kket{\Delta\,\treh} \,
 \bbra{\Delta\,\treh}I_{[1, 1, 1]}(Q^2)\kket{N\,\oneh}}
 {\sqrt{\omega'^2+m_{N}^2}-m_{\Delta}+\omega'}\right)_{\omega'= 0}.  \label{30}
 \ee Eqs.~(\ref{29}) and (\ref{30}) recover the results of Ref.~\cite{[Guichon]}
 obtained in a non-relativistic framework, while in the $u$-channel contribution
 the covariant formalism introduces a mixing between electric, Coulomb and
 magnetic multipole transitions, as a tipical relativistic effect due to the
 boost transformations from a frame to another one.

In Fig.  1, the curves corresponding to i) our calculation, obtained by using
the relativistic formalism presented in Sect.  2 and the $N-\Delta(1232)$ form
factors of Ref.~\cite{[CPSS]}; ii) the non-relativistic CQ model of
Ref.~\cite{[Guichon]}; and iii) the effective Lagrangian model of
Ref.~\cite{[VAN]} are compared.  The data point at the real-photon limit has
been obtained by taking the transition form factors extracted from the
experimental helicity amplitudes of PDG ~\cite{[PDG]} as input in our formalism.
On the one hand, it should be pointed out that such a data point, equal to
$(13.1 \pm 1.6)\times 10^{-4} $ fm $^3,$ is in fair agreement with the value
obtained in Ref.~\cite{[MNZ]} by using the experimental data on pion
photoproduction.  On the other hand, following the analysis of
Refs.~\cite{[NL],[Az]}, the disagreement with the theoretical calculations could
be ascribed to the lacking of a fully consistent treatment of the final state
interaction of the $N\pi$ system in the extraction of the experimental helicity
amplitudes.  The main difference between the model calculations of $\beta_{\rm
{para}}({q})$ stems from the different ${q}$ dependence of the transition form
factors combined with the relativistic effects due to the Lorentz
transformations.  The role of these different ingredients is disentangled in
Figs.  2 and 3, where the results of the non-relativistic CQ model of
Ref.~\cite{[Guichon]} have been considered as a reference, since such a model is
more close to our approach in terms of CQ.  In Fig.  2, the comparison with our
results, obtained as in Fig.  1, for the separate contributions from the direct
and crossed channels is shown.  In the direct term, since the Lorentz
transformations are equal to the identity, the different behaviour as a function
of ${q}$ is mainly given by the ${q}$ dependence of the transition form factors.
In particular, the effects due to the relativistic transition form factors are
more relevant at the real-photon point and for ${q}>0.9$ GeV/c, where it is
found a slower fall-off than the gaussian one predicted by the non-relativistic
CQ model.  The relativistic effects are more significant in the contribution of
the crossed term, where the action of the Lorentz transformations affects the
results mainly at increasing values of the momentum transfer.

A more detailed study of the effects of the Lorentz transformations can be done
with the aid of the results shown in Fig.  3, where the predictions of the fully
non-relativistic CQ model of Ref.~\cite{[Guichon]} are compared with the
calculations obtained with our relativistic formalism, Eqs.  (\ref{20}) and
(\ref{21}), but the same non-relativistic form factors as in the model of
Ref.~\cite{[Guichon]}.  As expected, the two calculations of the direct term
contribution overlap, whereas the discrepancy in the crossed channel
contribution gives a direct evidence of the importance of the Lorentz
transformations, whose effects grow for increasing values of ${q}$.

\section{Summary} We have investigated the nucleon generalized polarizabilities
within a relativistic framework.  In particular, we have formally derived
Lorentz covariant expressions for the reduced multipoles of the residual part of
the Compton tensor (see Eqs.  (\ref{20}) and (\ref{21})).  These multipoles
determine the generalized polarizabilities in the limit of vanishing real-photon
energy.  As a first application, we have calculated the $\Delta(1232)$
contribution to the nucleon paramagnetic polarizability, analyzing in details
the role of the relativistic formalism and the em transition form factors,
obtained within different models.  On the one hand, by working in the center of
mass of the final photon-nucleon system, the Lorentz covariance produces effects
only in the $u$ channel contribution, by introducing model-independent
kinematical corrections which become important at values of the three-momentum
transfer $q\ge 0.5$ GeV/c.  On the other hand, the $q$ evolution of the nucleon
paramagnetic polarizability is sizably different if one adopts the transition
form factors obtained within a relativistic (light-front) CQ model or within a
non-relativistic CQ model, even at low values of the momentum transfer.

\section{Acknowledgment} We are very grateful to S.  Boffi for suggesting the
topic investigated in this work and to A.  Metz and E.  Pace for helpful
discussions.

\section{Appendix}

In this Appendix the explicit form of the tensors $O^{K}_{k}$ and $T^{K}_{k}$
(see Eqs.~(\ref{20}) and (\ref{21})) is given.

In particular, the tensor $O^{K}_{k}(S,c;\vec p_X,\vec p_i,\vec p_f)$ is defined
in terms of the Wigner rotations $R=R_L(\vec p_X,\vec p_{i})$ and
$R'^{\dagger}=R^{\dagger}_L({\vec p}_X,{\vec p}_{f})$ (cf.  Eq.~(\ref{15}))
according to

\eq O^{K}_{k}(S,c;\vec p_X,\vec p_i,\vec p_f)=\sum_{\zeta,\gamma}
\,\cg{S}{\zeta}{c}{\gamma}{K}{k}
\mbox{Tr}\left\{R\,\sigma^S_{\zeta}\,R'^{\dagger}\,\sigma^c_{\gamma}\right\},
\label{31} \ee where the indices $S$ and $c$ can assume only two values,
($0,1$), and $\sigma^{S\,(c)}_{\zeta\,(\gamma)}$ are the spherical components of
the operator $\sigma_{\mu}=(I,\vec{\sigma}).$

The tensor $T^{K}_{k}$ is defined as \eqa T^K_k\left(\nu,\nu';\hat{ q},\hat {
p}_{\rm{o}i};\hat{ q}', \hat{ p}_{\rm{o}f};\frac{ \vec p_X}{m_X}\right)
=\sum_{\alpha,\alpha'}\, \cg{a}{\alpha}{a'}{\alpha'}{K}{k} {\Phi}^a_{\alpha}
\left(\varepsilon;\hat{ q},\hat { p}_{\rm{o}i};{ \vec p_X\over m_X}\right)\,
{\Phi}^{a'}_{\alpha'}\left(\varepsilon';\hat{ q}', \hat{p}_{\rm{o}f}; { \vec
p_X\over m_X}\right), \label{32} \medskip \eea where $\nu=(a,\varepsilon)$ and
$\nu'=(a',\varepsilon')$, while $\varepsilon$ and $\varepsilon'$ stand for the
set of indices $(b,s,\bar s;d,\ell,{\cal L})$ and $(b',s',\bar s',d',\ell',{\cal
L'})$, respectively.  In Eq.~(\ref{32}), the tensor ${\Phi}^a_{\alpha}$ is
explicitly given by \eq {\Phi}^a_{\alpha} \left(\varepsilon;\hat{q},
\hat{p}_{\rm{o}i};{\vec p_X\over m_X}\right)=\sum_{\delta,\beta}\,
\cg{d}{\delta}{b}{\beta}{a}{\alpha} \left\{ Y_{\ell}(\hat{q})\otimes Y_{{\cal
L}}(\hat{p}_{\rm{o}i})\right\}_{d,\delta} \Lambda^{b}_{\beta}\left(s, \bar s;{
\vec p_X\over m_X}\right), \label{33} \medskip \ee where the bipolar spherical
harmonics are defined by \eqa \left\{ Y_{\ell}(\hat q)\otimes Y_{{\cal L}}(\hat
p)\right\}_{d,\delta}= \sum_{m,\lambda}\, \cg{\ell}{m}{{\cal
L}}{\lambda}{d}{\delta} Y^{\ell}_{m}(\hat q) Y^{{\cal L}}_{\lambda}(\hat
p),\label{34} \eea and $\Lambda^{b}_{\beta}\left(s, \bar s;\vec p_X/ m_X\right)$
is given in terms of the spherical components of the Lorentz boost (see
Eq.~(\ref{13})) \eqa \Lambda^{b}_{\beta}\left(s, \bar s;{ \vec p_X\over
m_X}\right) &=&\sum_{m_s,\bar{m}_{s}}\, (-1)^{\bar{m}_{s}} \cg{s}{m_s}{\bar s}
{-\bar {m}_{s}}{b}{\beta} \,\Lambda^{s\,\bar s}_{m_s\,\bar {m}_{s}}\left (\frac{
\vec p_X}{m_X}\right).\label{35} \eea

\newpage \centerline{FIGURE CAPTIONS}

Figure 1:  The $\Delta(1232)$ contribution to the nucleon paramagnetic
polarizability vs the three-momentum transfer ${q}$.  Solid line:  results
obtained by using our relativistic formalism (cf.  Eqs.~(\ref{20}) and
(\ref{21})) and the transition form factors evaluated within a relativistic CQ
model~\cite{[CPSS]}; dotted line:  the non-relativistic CQ model of
Ref.~\cite{[Guichon]}; dot-dashed line:  the effective Lagrangian model of
Ref.~\cite{[VAN]}.  The data point at the real-photon point (slightly displaced
in order to show the error bar) is the result obtained from the experimental
helicity amplitudes of PDG~\cite{[PDG]} (see text).

Figure 2:  The direct and crossed term of the $\Delta(1232)$ contribution to the
nucleon paramagnetic polarizability vs the three-momentum transfer ${q}$.  The
solid and dashed lines are the direct and crossed term, respectively.  Thick
lines correspond to the predictions obtained by using our relativistic formalism
(cf.  Eqs.~(\ref{20}) and (\ref{21})) and the transition form factors evaluated
within a relativistic CQ model~\cite{[CPSS]}, while thin lines are the results
obtained within the non-relativistic CQ model of Ref.~\cite{[Guichon]}.

Figure 3:  The same as Fig.  2, but using in the covariant formalism the
transition form factors of the non-relativistic CQ model of
Ref.~\cite{[Guichon]}.  The results corresponding to the direct channel
contribution calculated by our relativistic formalism and by the fully
non-relativistic approach of Ref.~\cite{[Guichon]} overlap.

\newpage \begin{figure}\epsfxsize=15cm \epsfig{file=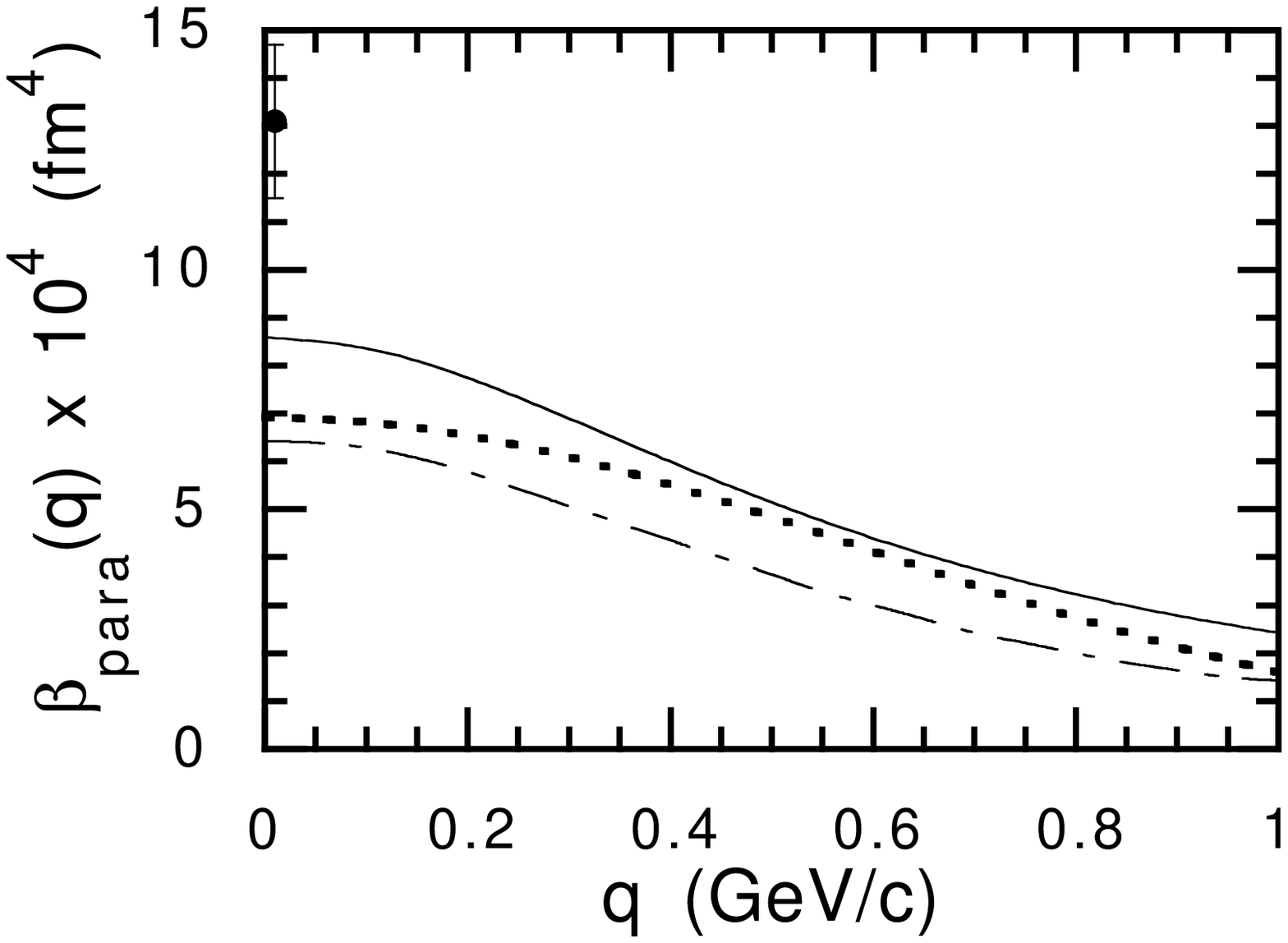}

\vspace {-1cm}

\centerline{Figure 1.  B.  PASQUINI and G.  SALM\`E} \end{figure} \newpage
\begin{figure}\epsfxsize=15cm \epsfig{file=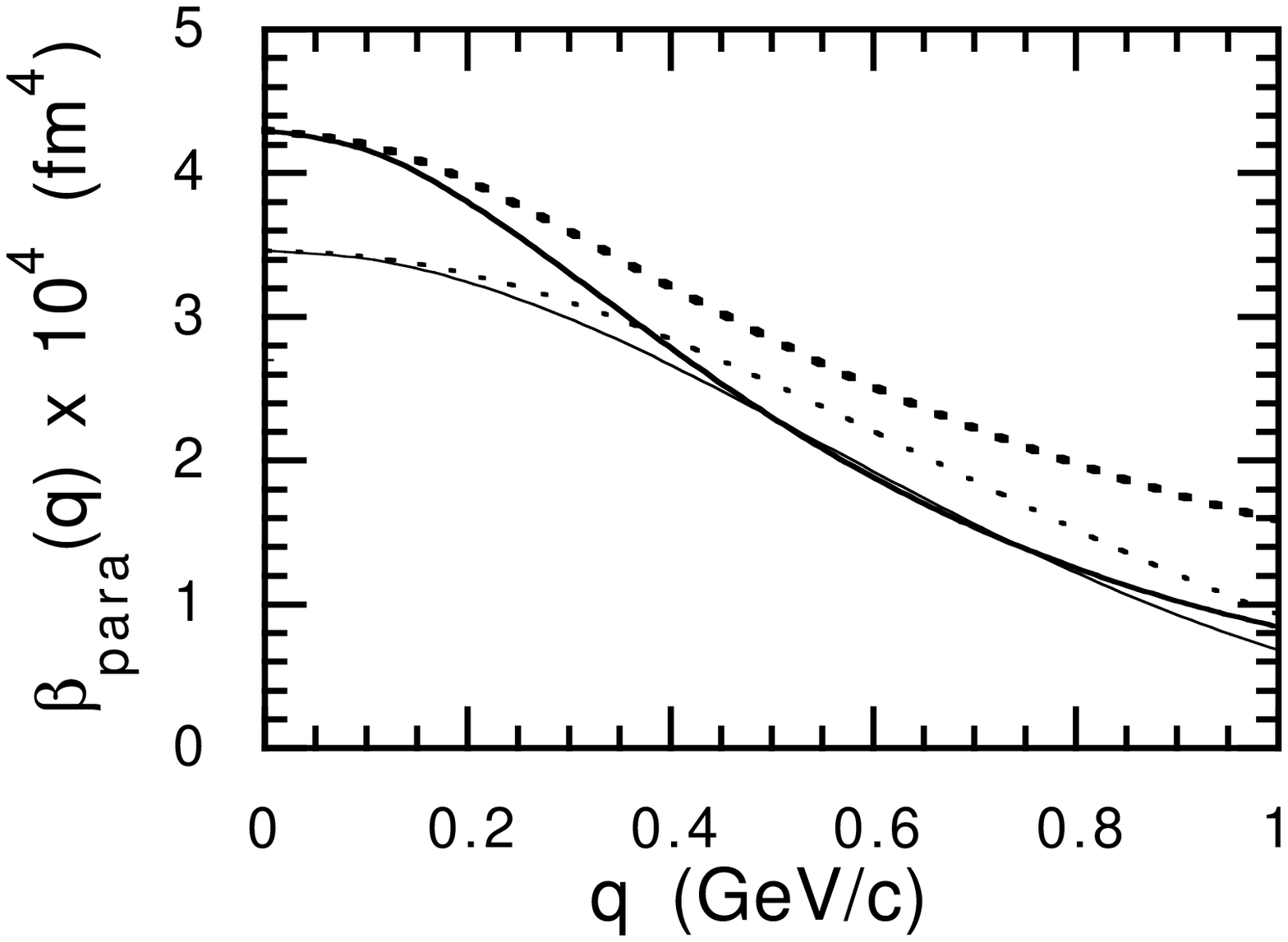}

\vspace {-1cm}

\centerline{Figure 2.  B.  PASQUINI and G.  SALM\`E} \end{figure}\newpage
\begin{figure}\epsfxsize=15cm \epsfig{file=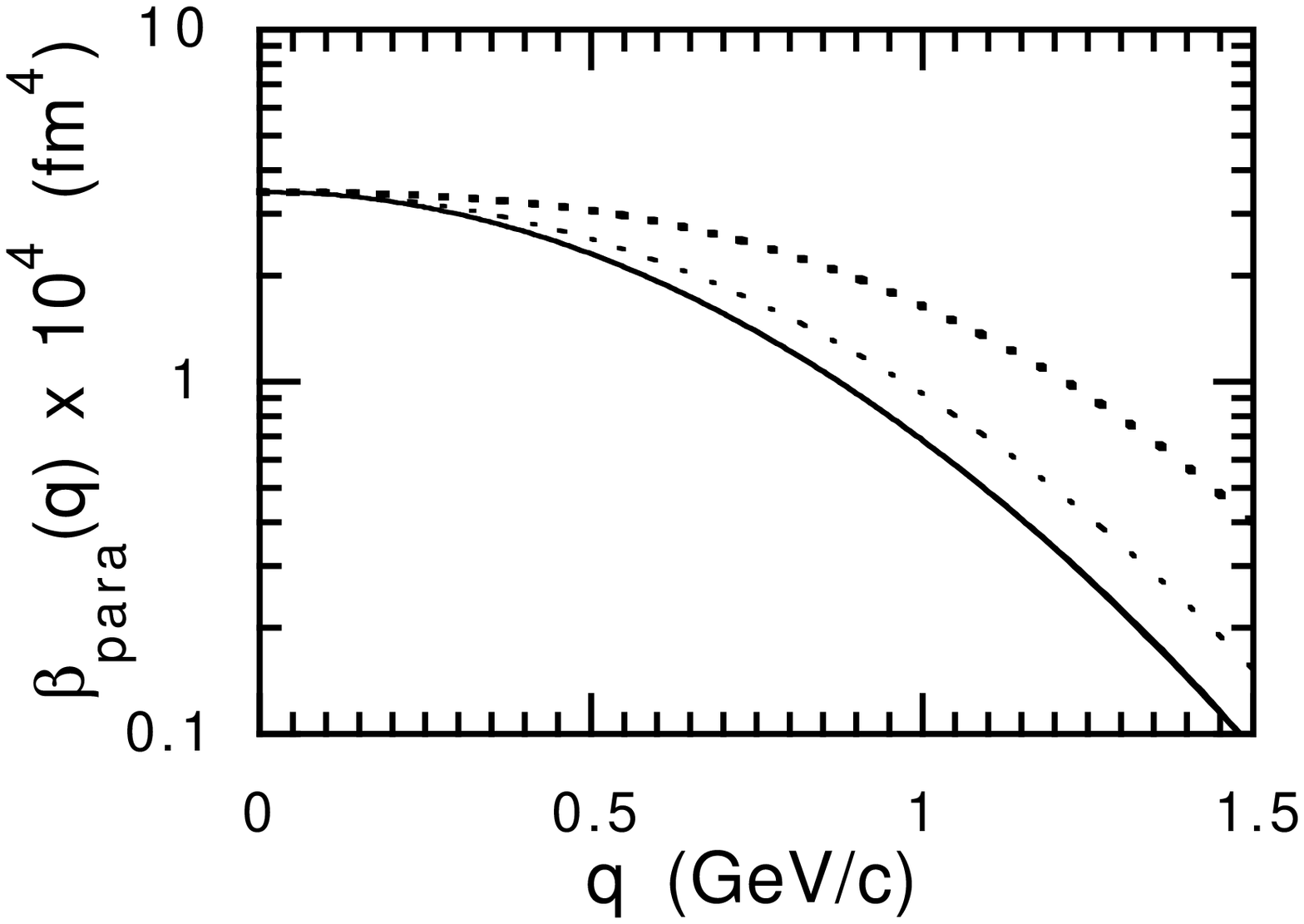}

\vspace {-1cm} \centerline{Figure 3.  B.  PASQUINI and G.  SALM\`E}\end{figure}
\end{document}